\documentstyle[sprocl,psfig]{article}

\def\beq{\begin{equation}}
\def\eeq{\end{equation}}
\def\beqn{\begin{eqnarray}}
\def\eeqn{\end{eqnarray}}
\def\nn{\nonumber\\}
\def\etal{{\it et al.}}
\def\Re{\,\mbox{Re}\,}
\def\Im{\,\mbox{Im}\,}

\begin{document}

\title{PRODUCTION AND DECAY OF ETA-MESIC NUCLEI}
\author{A. I. L'VOV}
\address{P. N. Lebedev Physical Institute, Russian Academy of Sciences
  \\ Leninsky Prospect 53, Moscow 117924, Russia }
\maketitle

\abstracts{
Using the Green function method, binding effects on produced
$\eta$-mesons in the two-stage reaction $\gamma + A \to N + \eta +
(A-1) \to N + (\pi N) + (A-2)$ are studied. The energy spectrum of the
correlated $\pi N$ pairs which arise from decays of $\eta$'s inside the
nucleus is strongly affected by an attractive $\eta$-nucleus optical
potential. Its resonant behavior gives a clear signal of formating
intermediate $\eta$-mesic nuclei.}

It was found long ago that the $S_{11}(1535)$ resonance, which lies
above the $\eta N$ threshold and is strongly coupled to the $\eta N$
channel, makes the low-energy $\eta N$ interaction attractive and leads
to an existence of bound $\eta$-nucleus systems, the so-called
$\eta$-mesic nuclei.~\cite{hai86}  This finding was later confirmed and
even strengthened.  With contemporary estimates of the $\eta N$
scattering length,~\cite{bati97,gree97}  the $\eta$-mesic nuclei
${}_\eta A$ are expected to exist for all $A \ge
3$.~\cite{raki96,trya97} Studies of the reactions like $p+d \to
{}^3\mbox{He} + \eta$ and $d+d \to {}^4\mbox{He} + \eta$ have already
provided an experimental evidence that the $\eta$ and the nucleus in
the final state experience a strong attraction which manifests itself
in a near-threshold enhancement and in a rapid energy dependence of the
cross section.~\cite{wilk93,will97}  Nevertheless, a direct observation
of bound rather than free etas would be more convincing for a discovery
of $\eta$-mesic nuclei. Since the bound $\eta$ eventually decays
through the subprocess $\eta N\to \pi N$, a clear signal for a presence
of the stopped etas in nuclei would be in an observation of final pions
and nucleons with almost opposite momenta, with the kinetic energies of
about 300 MeV and 100 MeV, respectively, and with the total energy
close to $m_\eta + m_N$.~\cite{soko91}  In the present work, production
of such pairs is studied within a simple model which is aimed at
learning how the attraction between the eta and the nucleus affects
characteristics of the pairs.

In accordance with the original suggestion,~\cite{soko91}  we consider
the two-stage reaction
\beq
\label{reaction}
   \gamma + A \to N_1 + \eta + (A-1) \to N_1 + (\pi N_2) + (A-2),
\eeq
in which the fast nucleon $N_1$ knocked out in the subprocess
\beq
\label{stage1}
  \gamma(k) + N_\alpha \to \eta(E_\eta) + N_1
\eeq
escapes from the nucleus, whereas the $\eta$ collides with another
nucleon $N_\beta$ in the nucleus and perishes producing a pair which
also escapes:
\beq
\label{stage2}
  \eta + N_\beta \to \pi + N_2.
\eeq
Considering the rest of the nucleus as frozen, we write the matrix
element of (\ref{reaction}) as
\beqn
\label{amplitude}
   T_{\alpha\beta} &=&
   F_\gamma(k) F_\eta(E_\eta)
   \int \hspace*{-1.5ex} \int  e^{i\vec k \vec r_1}
   \psi_\alpha(\vec r_1) \, \psi_{N_1}^{(-)*}(\vec r_1)
\times \nn && \hspace*{-2em}
   \psi_\beta(\vec r_2) \, \psi_\pi^{(-)*}(\vec r_2) \,
   \psi_{N_2}^{(-)*}(\vec r_2) \,
    G(\vec r_1, \vec r_2; E_\eta) \,d\vec r_1 \, d\vec r_2.
\eeqn
Here $\psi_\alpha$, $\psi_\beta$ are the wave functions of the bound
nucleons with the binding energies $\epsilon_\alpha$, $\epsilon_\beta$
and $\psi_{N_1}$, $\psi_{N_2}$, $\psi_\pi$ are the wave functions of
the final particles.  $F_\gamma$, $F_\eta$ are the amplitudes of the
reactions (\ref{stage1}) and (\ref{stage2}) which, at energies
considered, are approximated by $s$-waves. The Green function $G$ gives
the amplitude of $\eta$ with the energy
\beq
  E_\eta = E_\gamma + \epsilon_\alpha - E_{N_1}^{\rm kin}
    = E_\pi + E_{N_2}^{\rm kin} - \epsilon_\beta
\eeq
to propagate from $\vec r_1$ to $\vec r_2$ in the mean field $V(r)$ of
the intermediate nucleus $(A-1)$ which can be assumed to be independent
on $\alpha$. In the following we also neglect the dependence of
$E_\eta$ on the hole states $\alpha$, $\beta$ and replace
$\epsilon_\alpha$, $\epsilon_\beta$ by their Fermi-gas average
$\langle\epsilon\rangle \simeq -23$~MeV.

When a single bound state $\psi_0$ of the (complex) energy $E_0$
dominates, the Green function takes the separable form
\beq
\label{factorize}
  G(\vec r_1, \vec r_2; E_\eta) \simeq
   \frac{\psi_0^+(\vec r_1) \, \psi_0(\vec r_2)} {E_\eta^2 - E_0^2},
\eeq
which results in the Breit-Wigner resonant behavior of the pair
production through the intermediate $\eta$-mesic nucleus.  In such an
approximation, the amplitude (\ref{amplitude}) depends on the overlap
of $\psi_0(r)$ with the nucleus's nucleons and typically the total
cross section of the $\eta$-mesic nucleus formation by photons is 5--10
$\mu$b for $A=12$ to 16.~\cite{trya95}  With the realistic optical
potential strength, however, there are several bound states of $\eta$
which are strongly overlapped and act coherently.  Also, there is a
non-resonance background which describes the process $\gamma \to \eta
\to \pi$ in the nucleus with unbound etas.  For these reasons
Eq.~(\ref{factorize}) is generally insufficient and the full Green
function has to be used to describe the reaction (\ref{reaction}).

As an illustration of what may happen, we discuss here the spectral
function~\cite{mori85}
\beq
   S(E_\eta) = \int \hspace{-1.5ex} \int \rho(\vec r_1)\,\rho(\vec r_2)
    \, |G(\vec r_1, \vec r_2; E_\eta)|^2  \, d\vec r_1 \, d\vec r_2,
\eeq
which describes a global nuclear dependence of the matrix element
(\ref{amplitude}) squared and averaged over the nuclear states and
momenta of the outgoing particles. $S(E_\eta)$ characterizes the
nuclear dependence of the total cross section of the two-step
transition $\gamma\to\eta\to\pi$ in nuclei.  It is proportional to the
number of nucleons hit by $\eta$'s produced somewhere inside the
nucleus. This number increases when the $\eta$ has the energy close to
a resonance level; such $\eta$'s are captured by the nucleus and pass a
few times across the nucleus before they annihilate or escape.

In actual calculations of $G$ and $S(E_\eta)$ we use the simple
first-order energy-dependent potential $2E_\eta\,V(r,E_\eta) = -(4\pi
\sqrt{s}/M_N) f_{\eta N}(E_\eta)\,\rho(r)$ with the $\eta N$ scattering
amplitude taken from Ref.~\cite{gree97} and with the square-well
nuclear density $\rho(r)=0.75 \rho_0$ at $r<R_A$, $R_A = 1.2
A^{1/3}$~fm.  Such a potential gives the energy of the ground state and
its width $\Gamma$ close to those found in a recent
analysis~\cite{kulp98}.  Typically, the widths are $\Gamma \sim 20$ MeV
and far less than those found in an older work~\cite{oset88} which
seems to overestimate~\cite{kulp98} the width's broadening due to the
two-nucleon absorption $\eta NN \to NN$ in the nucleus.

\begin{figure}[htb]

\vspace*{-10.2cm} \hspace*{-0.5cm}
\psfig{file=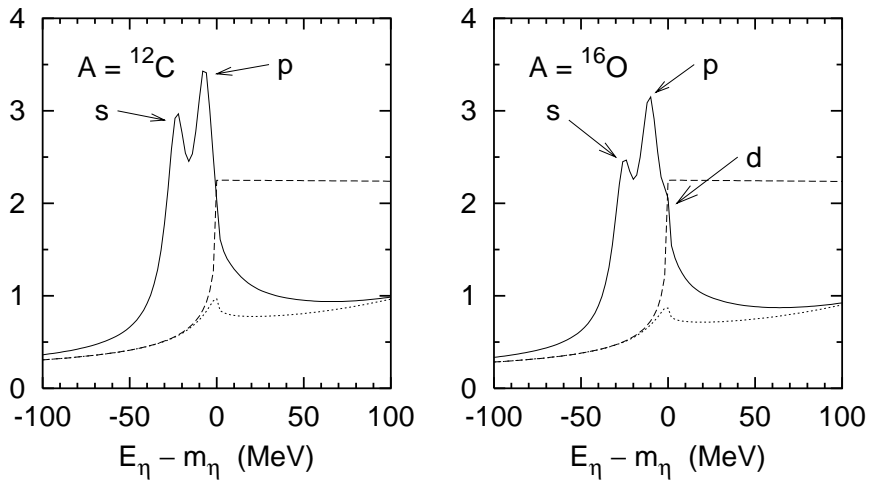,width=12cm}

\caption{The normalized spectral function $\bar S(E_\eta) = (16\pi^2
R_A^2 /A^2) S(E_\eta)$ for the square-well potential representing the
carbon and oxygen. Dashed lines: the optical potential is switched off;
then $\bar S(E_\eta) = \frac94$ above the $\eta$ threshold. Dotted
lines:  the absorption $\Im V(r)$ is on.  Solid lines: both the
attraction $\Re V(r)$ and the absorption $\Im V(r)$ are on.  The
resonance-like structures are composed of $s$, $p$, $d$ resonances in
the $\eta$-nucleus system.}

\vspace*{-0.2cm}
\end{figure}

In the absence of the potential $V(r)$, the Green function reads $G =
e^{iqr}/(4\pi r)$, where $r=|\vec r_1 - \vec r_2|$ and $q^2 = E_\eta^2
- m_\eta^2$. Accordingly, $S(E_\eta)$ does not depend on $E_\eta$ when
$E_\eta > m_\eta$.  At sub-threshold energies, when $\eta$ cannot
propagate far from the production point, $S(E_\eta)$ rapidly vanishes.
When the absorptive part $\Im V(r)$ of the optical potential is taken
into account, $S(E_\eta)$ falls down as well.  However, it strongly
enhances when the attraction $\Re V(r)$ is on and the bound states
appear.  In fact, the resonance-like structure of $S(E_\eta)$ consists
of many $s$, $p$, $d$, \ldots wave contributions.  See Fig.~1.

The practically important finding is that the non-resonance background
in $S(E_\eta)$ is relatively small, so that most of produced $\pi N$
pairs with near-threshold energies appear from the decay of the
resonant $\eta$-mesic states.  Due to the spread in the separation
energies $\epsilon_\alpha$, $\epsilon_\beta$, the inclusive
distribution of the total energy $E_\pi+E_{N_2} = E_\eta + m_N +
\epsilon_\beta$ of the pairs is smeared and rather exhibits a single
giant peak of the width $\Gamma \sim 40{-}50$ MeV.  Such pairs have
been recently observed in the experiment performed at Lebedev
Institute.~\cite{LPI98}  Further analysis of their energy distribution
may hopefully reveal whether the $\eta$-mesic nuclei were really found.

This research was supported in part by the Russian Foundation for Basic
Research, grant 96-02-17103. Useful discussions with G.A.~Sokol are
highly appreciated.

\end{document}